\documentclass[12pt]{article}

\usepackage{arxiv}
\usepackage{times}
\usepackage{epsfig}
\usepackage{graphicx}
\usepackage{float}
\usepackage{amsmath}
\usepackage{amssymb}
\usepackage{color}
\usepackage[table]{xcolor}
\usepackage{multirow}
\usepackage{subcaption}
\usepackage{makecell}
\usepackage{enumitem}

\usepackage[pagebackref=true,breaklinks=true,colorlinks,bookmarks=false]{hyperref}

\graphicspath{{images/}}
\DeclareGraphicsExtensions{.eps,.png,.pdf,.jpg,.jpeg,.JPG}

\title{Residual Feature Pyramid Network for Enhancement of Vascular Patterns}

\author{
Ketan~Kotwal\hspace{1mm}\mbox{\href{https://orcid.org/0000-0003-3766-0881}{\includegraphics[width=4mm]{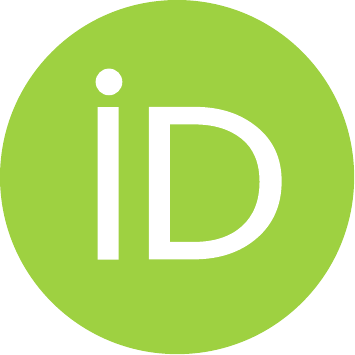}}}\\
Idiap Research Institute, Switzerland\\
\texttt{ketan.kotwal@idiap.ch}\\
\And
S\'{e}bastien Marcel\hspace{1mm}\mbox{\href{https://orcid.org/0000-0002-2497-9140}{\includegraphics[width=4mm]{images/orcid.pdf}}}\\
Idiap Research Institute, Switzerland\\
University of Lausanne, Switzerland\\
\texttt{ sebastien.marcel@idiap.ch}\\
}
\date{}

\hypersetup{
pdftitle={Enhancement of Vascular Patterns},
pdfsubject={},
pdfauthor={Ketan Kotwal},
pdfkeywords={Biometrics, Vein Recognition, Residual FPN},
}

\begin{document}
\maketitle

\begin{abstract}
The accuracy of finger vein recognition systems gets degraded due to low
and uneven contrast between veins and surroundings, often resulting in
poor detection of vein patterns. We propose a finger-vein enhancement
technique, ResFPN (\textit{Residual Feature Pyramid Network}), as a generic
preprocessing method agnostic to the recognition pipeline. A bottom-up
pyramidal architecture using the novel Structure Detection block
(SDBlock) facilitates extraction of veins of varied widths. Using a
feature aggregation module (FAM), we combine these vein-structures, and
train the proposed ResFPN for detection of veins across scales.  With
enhanced presentations, our experiments indicate a reduction upto 5\% in
the average recognition errors for commonly used recognition pipeline
over two publicly available datasets. These improvements are persistent
even in cross-dataset scenario where the dataset used to train the
ResFPN is different from the one used for recognition.
\end{abstract}

\section{Introduction}
\label{sec:intro}
\begin{figure}[t]
\begin{center}
\includegraphics[width=\columnwidth]{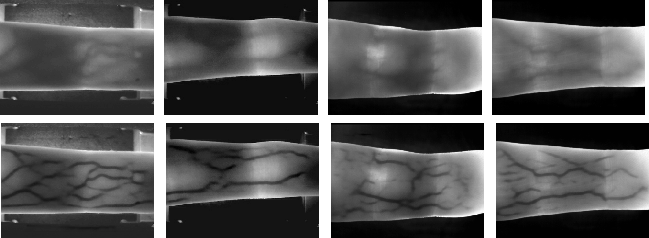}
\end{center}
\caption{The top row shows examples of original (acquired by sensor) FV presentations: two each from SDUMLA~\cite{sdumla} and UTFVP~\cite{utfvp} datasets. 
The corresponding images from the bottom row are the results of the proposed vein enhancement technique.}
\label{fig:example1}
\end{figure}

Use of vascular patterns as the biometric recognition trait is becoming more
prevalent due to its distinctive advantages such as high recognition accuracy,
difficulty in spoofing, and less interference of external factors. Typically,
veins of finger(s), palm, and wrist are popular biometric modalities. In this
work, we consider only finger vein (FV) as the biometric modality. The
reflection-based FV scanners can be constructed in a contactless manner---which
makes them an attractive biometric modality offering a better user experience
and alleviating hygiene concerns (that may occur in enclosure or touch-based
vein scanners). The performance of FV recognition pipeline is strongly
correlated to the quality of the FV presentation acquired by the near-infrared
(NIR) sensor (\textit{i.e.} camera). These blood vessels lie beneath the skin of the
subject and therefore do not always appear prominent in the acquired
presentation. Figure~\ref{fig:example1} shows (see top row) some samples of FV
presentations where the vein structures are not clearly visible across the
region.  Due to lack of contrast and uneven illumination, these presentations
often suffer from poor feature extraction, and subsequently result in low and
incorrect matching scores impeding the performance of the overall FV recognition
system. In this work, we propose a deep learning (DL)-based technique for
enhancement of vein structures in the presentations acquired in the NIR spectra.
The proposed technique is independent module that can be plugged into an
existing FV recognition pipeline at the preprocessing stage. 

An overall FV recognition pipeline can be built from conventional image
processing techniques or it can be based on a deep convolutional neural network
(CNN). Typically, in both cases, the NIR presentation is first preprocessed for
cropping, resizing, and orientation correction. The conventional processing
pipeline employs feature extraction block to generate a feature descriptor (it
acts as reference or template for enrolment data), followed by the matching
block that computes similarity metric between feature descriptor of the test
sample (also known as probe) and predefined templates. The DL-based FV
recognition pipelines usually combine both blocks by modeling the recognition
task as an $n$-class classification problem. A cascade of convolutional and
pooling layers learns vein-related features which are then transformed into
class probabilities by one or more fully connected layers. For any pipeline,
conventional or DL-based, efficient extraction or learning of relevant features
from input presentations is the key to build a highly accurate recognition
system. Popular feature extraction methods such as repeated line
tracking~\cite{rlt}, wide line detector~\cite{wld}, and maximum curvature
(MC)~\cite{mc} are based on computation of the local gradient or cross-sectional
profile of pixels as the first step. The efficacy of these quantities (gradient
or profile) is directly proportional to the contrast in the image. The deep
CNNs, as well, are susceptible to distortion in the quality of input images such
as noise, blur, and contrast~\cite{dodge2016, hossein2017, dodge2017}. This
essentially reinforces the importance of good contrast (between vein structures
and surroundings) in designing a highly accurate FV recognition pipeline. It
may be noted that the publicly available FV datasets are relatively much smaller
(in the range of 2000--3000 total presentations), furthermore, only a fraction
of entire dataset is used for training purposes. Since training deep CNNs with
small amount of data is challenging, improving the quality of the input
presentations-- by enhancing the vein structures-- can be of significant
assistance in training as well as inference. From the existing literature, it
appears that the problem of enhancement of vein structures, particularly using
learning-based methods, has not received much attention despite its apparent
usefulness.

Using the presentations as captured by the NIR sensor without the aforementioned
enhancement has two serious shortcomings: (1) Due to variable width of blood
vessels and variable local contrast (because of presence of tissues around
vessels), the feature extraction may detect fewer vein-structures from the
presentation. The subtle vein structures-- that may carry subject-specific
discriminatory information-- may remain undetected. Alternatively, one has to
extensively experiment with parameters of feature extractor or CNN to obtain
good recognition accuracy. (2) Since the parameters of pipeline are tuned for
specific dataset or sensor, the FV recognition system can suffer from poor
generalization across different datasets. In case of change of NIR sensor, which
is a quite common real life use-case, one has to rely on expensive and
time-consuming solution of capturing new dataset to tune the parameters or train
the CNN.

To address these concerns, we propose a deep CNN-based method for enhancement of
vein structures from the FV presentations acquired in NIR channel. Our network
accepts an NIR presentation in the form of single channel image; and generates
an image consisting of vein-\textit{like} structures. This result is combined
with the input to obtain the enhanced presentation which can then be processed
by any FV recognition pipeline. Samples of enhanced images obtained from our
work are shown in (the bottom row of) Figure~\ref{fig:example1} where the
appearance of veins is much sharper, clearer, and visibly darker as compared to
their unprocessed/ original versions. With good contrast around veins, these
presentations are less sensitive to the parameters of feature extraction method
or model. We train our model using the vein annotations (manually generated
binary labels) as the target.

As vein structures exhibit variable width or thickness, the choice of spatial
resolution (or scale) is crucial in designing the enhancement network. Our
network consists of structure detection blocks at multiple resolutions akin to
the feature pyramid networks (FPNs)\cite{fpn2017}. The vein structures (or their
parts) detected at each level are combined through a feature aggregation module
(FAM) to get a fused output. We design a structure detection block (SDBlock) as
the basic unit of our network---that detects vein structures and also generates
a set of feature maps, at reduced resolution, for processing by subsequent
blocks. Through residual architecture, our network is able to extract FV
structures across scales and fuse those to obtain an enhanced FV presentation.
The contributions of our work can be summarized as follows:
\begin{itemize}[noitemsep]
\item We have designed a fully convolutional Residual FPN (\textbf{ResFPN}) for enhancement of vein structures.
This architecture, consisting of only 600$k$ parameters, efficiently detects vein structures of varied thickness without
need for any specific tuning.
\item We have introduced a novel unit for structure detection, SDBlock. Through the SDBlock, we are
able to achieve two objectives simultaneously: extraction of vein structures and generation of input for next layers/blocks.
\item Through indirect assessment of work, we demonstrate the efficacy of the proposed enhancement
technique: the average error rate of FV recognition performance on publicly available datasets reduced
upto 5\% after enhancing the presentations by ResFPN. This improvement has been validated in intra- and
cross-dataset testing scenarios.  
\end{itemize}

In Section~\ref{sec:rw}, we briefly describe existing works related to FV
enhancement. The proposed ResFPN is described in Section~\ref{sec:method}. We
provide experimental results along with details of datasets and evaluations
measures in Section~\ref{sec:results}. Finally, Section~\ref{sec:conc} provides
concluding remarks.


\section{Related Work}
\label{sec:rw}

Kumar and Zhou~\cite{kumar2012} generated an average background image for a
sub-block of input FV presentation, followed by local histogram equalization. A
combination of edge preserving filtering, elliptic highpass filtering and
histogram equalization was proposed by Pi~\textit{et al.}~\cite{pi2010}. The contrast
limited adaptive histogram equalization (CLAHE) has been considered towards
enhancement of vein region by several works~\cite{kumar2012, banerjee2018,
kauba2014}. The use of Gabor wavelets at various scales and orientations for
enhancement of venous regions has been proposed by Yang and
Shi~\cite{yang2014}. They have also devised a scattering removal method for
better visibility of the acquired presentation. Methods in~\cite{shin2014,
pham2015} also advocate the use of Gabor filters for enhancement of FV
presentations.

Peng~\textit{et al.} proposed a non-local means (NLM)-based technique for enhancement of
veins in the NIR presentation~\cite{peng2013}. Their work is based on the
availability of several local patches with similar vein structures. These
multiple patches have been exploited to enhance the vein-structures. A recent
work by Zhang~\textit{et al.} combines the guided filter and tri-Gaussian model for FV
image enhancement~\cite{zhang2021}.

All aforementioned approaches for enhancement of vein regions are based on
conventional image processing techniques. Despite success of deep CNNs in
enhancement or restoration of images, very few works have studied DL-based
approaches for this task. In~\cite{qin2017}, a fully convolutional network
(FCN) has been developed to enhance the vein patterns, more specifically to
recover the missing segments within vein patterns. The training data were
created by randomly cropping some pixels from the FV images, and the
corresponding FCN was trained using MSE (mean square error) loss between the
output of the FCN and original image. Recently, Bros~\textit{et al.} proposed a deep
autoencoder-based method for enhancement of FV presentations~\cite{biosig2021}.
They used presentations enhanced with vein-annotations to train their network
by reducing the MSE loss.

\section{ResFPN for Vein Enhancement}
\label{sec:method}

In this section, we first describe the architecture of the proposed ResFPN for
FV enhancement along with our rationale in designing its building blocks.
Subsequently we provide details of training procedure and formulation of loss
function.

\begin{figure*}[t]
\centering
\begin{subfigure}[b]{0.55\textwidth}
    \includegraphics[width=\textwidth]{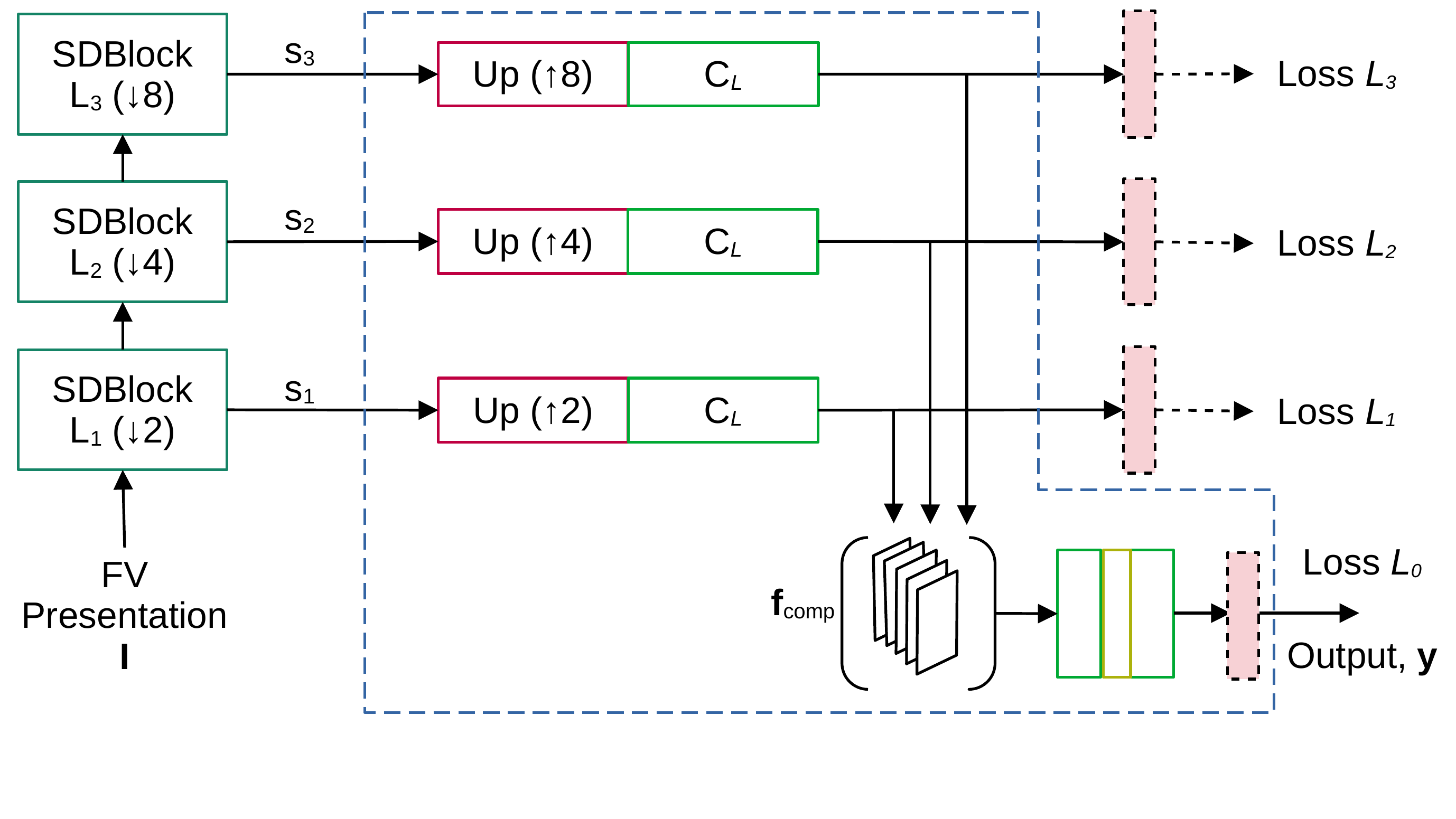}
    \caption{}
    \label{fig:resfpn}
\end{subfigure} \,
\begin{subfigure}[b]{0.35\textwidth}
    \includegraphics[width=\textwidth]{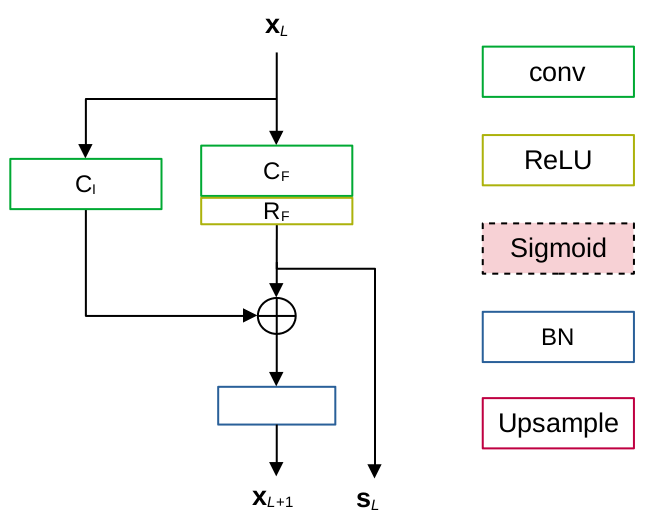}
    \caption{}
    \label{fig:sdblock}
\end{subfigure}
\caption{Architectures of the proposed finger vein enhancement technique: (a) ResFPN and (b) SDBlock. The blue dotted lines in (a) represent FAM.}
\label{fig:method}
\end{figure*}

\subsection{Network Architecture} 

Learning features at all scales from a combination of bottom-up pathway,
top-down pathway, and lateral connections-- also known as feature pyramid
network (FPN)-- has been shown to be efficient generic feature
extractor~\cite{fpn2017, wang2020}. When analyzed locally, vein pattern is a
structure of variable thickness; and extracting such a structure would require
learning a set of convolutional filters at different spatial resolutions or
scales. Based on the idea of FPN~\cite{fpn2017}, we construct a multi-level
bottom-up pathway to extract vein features at different scales.
Figure~\ref{fig:resfpn} shows the overall architecture of the proposed ResFPN.
We call each unit of this pathway as the \textsl{structure detection block}
(\texttt{SDBlock})---which will be described later in this section. At each
successive level, the SDBlock extracts vein-\textit{like} structures from
larger receptive fields as the spatial dimensions (resolution) of the
corresponding feature maps gradually reduce. Vein structures so-obtained from
each SDBlock are then combined by a feature aggregation module (\texttt{FAM})
which normalizes them in terms of resolution and number of feature maps. As
each SDBlock extracts only a part of overall vein pattern (depending on its
width or thickness), we combine the normalized outputs of FAM into a single
channel representation of the extracted structures. The vein-enhanced
presentation is obtained by a linear combination of the output of our network
and the original input presentation. \\

\noindent\textbf{Structure Detection Block (SDBlock):}

The architecture of SDBlock-- the fundamental unit for extraction of vein-like
structures-- is shown in Figure~\ref{fig:sdblock}. Detection of thin and
subtle vein-structures is accomplished by learning a set of convolutional
kernels, followed by a non-linear activation such as ReLU. (In
Figure~\ref{fig:sdblock} corresponding convolutional and ReLU layers are
represented as $C_F$ and $R_F$, respectively.) The size of kernels can be
calculated by analyzing the nominal width of vein structures at original
resolution. We employ a stride of 2 across the $C_F$ layer which implicitly
reduces the spatial dimensions, while no explicit spatial-level feature pooling
is used. The outputs of $R_F$ are structure features ($\mathbf{s}_L$) extracted
by the $L$-th SDBlock. 

Given the nature of vein structures, the feature extraction process is akin to
learning a set of \textit{bandpass} filters. The output of such filters
strongly suppresses or removes the information content beyond their effective
bandwidth. Therefore, using these outputs ($\mathbf{s}_L$) directly for
detection of structures (that are predominantly present in the possibly
suppressed frequency bands) is likely to render ineffective results.
Therefore, we propose to implement the shortcut (using the ResNet terminology)
by adding the input of the SDBlock to the output of $R_F$. The input is passed
through a convolutional layer $C_I$ to align its dimensions (spatial dimensions
and number of feature maps) to those of structure features, $\mathbf{s}_L$.
After the addition of shortcut, the corresponding output ($\mathbf{x}_{L+1}$)
is normalized at batch level ($\mathrm{BN}$) which may then be fed to the next
SDBlock.

If $\mathbf{x}_L$ is the input to the SDBlock, which could be the FV
presentation ($\mathbf{I}$) or feature maps generated by previous SDBlock, the
functioning of the $L$-th SDBlock is summarized below.

\begin{align}
\label{eq:sdblock}
    &\mathbf{s}_L = R_F \big( C_F (\mathbf{x}_L) \big) \\ 
    &\mathbf{x}_{L+1} = \mathrm{BN} \big( C_I(\mathbf{x}_L) + \mathbf{s}_L  \big)
\end{align}

The SDBlock, thus, accepts feature maps (or the input presentation), and
generates two outputs: (1) the residual structure features-- to be processed by
the FAM for output, and (2) normalized feature maps-- to be processed by the
SDBlock at next level. The process of detection of $\mathbf{s}_L$ features
operates in different frequency bands for each SDBlock. The proposed
architecture simplifies these objectives using shortcut connections: the
residual component is trained to learn structures in the feature maps; while
the combined/ summed component, boosted with detected features, is suitable for
similar processing at the next scale.\\

\noindent\textbf{Feature Aggregation Module (FAM):} 

The FAM receives structure features, $\mathbf{s}_L$, from each SDBlock; and as
the first step normalizes them through upsampling and $1 \times 1$
convolutions. For each SDBlock, the structure features are computed on
successively reduced scale (spatial resolution) of feature maps. We use nearest
neighbor-based interpolation to upsample the structure features to the scale of
original input presentations. Thus, no learning parameters are involved at
upsampling stage. Using $1 \times 1$ convolutions, we convert the feature maps
of each SDBlock into the same channel-dimension, say $n_\mathrm{ch}$, and refer
to them as $\mathbf{\widehat{s}}_L$.

As each $\mathbf{\widehat{s}}_L$ is upsampled to the resolution of input
presentation, the upsampling factor of $\mathtt{up}_L$ is determined
accordingly. If the network consists of $k$ SDBlocks, we obtain a composite
feature map with $k \, n_\mathrm{ch}$ channels whose spatial dimensions are
same as that of the input presentation. This composite feature map,
$\mathbf{f}_\mathrm{comp}$, represents the aggregation of vein features learnt
across multiple scales. We fuse the individual feature maps of
$\mathbf{f}_\mathrm{comp}$ into a single channel output,
$\mathbf{\widehat{y}}$, using two layers of convolutional layers with an
intermittent ReLU activation. The final output, $\mathbf{y}$, is obtained
through sigmoidal activation of $\mathbf{\widehat{y}}$.

The functions of the FAM are summarized below.
\begin{align}
    &\mathbf{\widehat{s}}_L = C_\mathrm{L} \big( \mathtt{up}_L (\mathbf{s}_L ) \big)\\
    &\widehat{\mathbf{y}} = C_2 \Big( R_1 \big( C_1(\mathtt{concat} \{ \mathbf{\hat{s}}_L \})  \big) \Big)\\ 
    &\mathbf{y} = \mathtt{Sigmoid}(\widehat{\mathbf{y}}) \nonumber
\end{align}

The enhanced presentation, $\mathbf{I}_E$, is obtained by linear combination of
the output, $\mathbf{y}$, and the input presentation, $\mathbf{I}$ using a
predefined weight $\alpha \in (0, 1)$ as $\mathbf{I}_E = \alpha \mathbf{y} + (1
- \alpha) \mathbf{I}$.

\subsection{Loss Function}

We formulate the problem of detection of vein structures as a binary
classification problem that assigns a probability of being a (part) of vein to
each pixel. The loss function, therefore, is defined as the binary cross
entropy (BCE) between the vein-annotations (a binary image with vein marking)
and the output, $\mathbf{y}$, of the ResFPN. The outputs of each SDBlock post
dimensional normalization, are expected to have extracted parts of vein
pattern. Therefore, we also propose to calculate loss over each of normalized
feature maps, $\mathbf{\widehat{s}}_L$. These feature maps are passed through a
sigmoidal activation, the BCE loss is computed for each of $n_\textrm{ch}$
feature maps, and then averaged to yield a scalar value. The overall loss
function, $\mathcal{L}$, is defined as summation of losses computed over each
of $L$ levels, and the loss computed on final output. If we denote the
vein-annotations as $\mathbf{y}_\mathrm{target}$, then the expression for
overall loss function is provided by Equation~\ref{eq:loss}.

\begin{equation}
\label{eq:loss}
    \mathcal{L} = \mathcal{L}_\mathrm{BCE}(\mathbf{y}_\mathrm{target}, \, \mathbf{y}) + 
    \sum_{k=1}^{L}  \mathcal{L}_\mathrm{BCE}(\mathbf{y}_\mathrm{target}, \, \mathtt{Sigmoid}(\mathbf{\widehat{s}}_k))
\end{equation}

\subsection{Training Procedure}
A small size of vein dataset and cumbersome task of manual annotation of vein
structures drastically limit the scope of training large deep networks. In
addition to designing a deep network with relatively fewer parameters, we have
incorporated data augmentation by flipping it along horizontal and vertical
axes. Each input presentation generates 4 samples (2 by horizontal flip and 2 by
vertical flip) which are then shuffled during training. Note that each
presentation is flipped to create $4 \times$ data contrary to typical
augmentation strategies where either original or flipped data are considered
(flipping takes place randomly). The input presentations are rescaled to a fixed
size ($320 \times 240$ in our case) to ensure consistency across different
datasets. The vein-annotations, acting as targets, were also processed in the
same manner. For training the ResFPN, we have chosen the Adam optimizer with a
learning rate of 1.0e-4. To generate the enhanced presentation, we have used 
$\alpha = 0.10$ to combine the vein-structures with input.


\section{Experiments and Results}
\label{sec:results}

We begin this section with details of the FV datasets and the protocols
designed for our experiments. Since there are no direct methods to access the
performance of enhancement, we have considered indirect assessment of our work
by measuring the difference in the performance of overall FV recognition
without and with application of our enhancement technique. We employ a
conventional FV recognition pipeline that consists of preprocessing functions
(cropping, orientation correction, resizing, etc.), followed feature extraction
using Maximum Curvature (MC) technique~\cite{mc}. The Miura Matching
technique~\cite{rlt} is used to compute the similarity or matching score
between the probe and model. We calculate the performance of FV recognition
using the measures described in Section~\ref{ss:measures}. Then we have
provided results of our experiments on conventional FV recognition pipeline.
The python code to reproduce the experimental results will be released
publicly.\footnote{Repository of Python code for experiments described in this
work: \url{https://gitlab.idiap.ch/bob/bob.paper.resfpn_cvprw}}

\subsection{Datasets and Protocols}

For experiments, we have used two publicly available datasets:
SDUMLA~\cite{sdumla} and UTFVP~\cite{utfvp}. The SDUMLA dataset consists of FV
images of 6 fingers (3 finger of each hand) from 106 individuals. This
collection has been repeated 6 times (called as sessions) to obtain a total of
3,816 FV presentations with $320 \times 240$ pixels in size. As we consider
each finger as a separate entity for our experiments, the SDUMLA dataset is
considered to have $106 \times 6 = 636$ clients. It should be also noted that
vein annotations are available only for session-I. We require this dataset for
two tasks: (1) to train and validate the ResFPN for enhancement; and (2) to
validate the overall FV recognition pipeline. The first task requires a split
of presentations to train the CNN, and to validate its performance over
training epochs. The second task requires two disjoint sets of data: one to
obtain score-related thresholds (\texttt{dev}), and another to evaluate the
performance of FV recognition using these score thresholds (\texttt{eval}). In
each subset, a further split of samples is required to enroll (\textit{i.e.}, to build
models), and samples to probe. We have created a \textit{Nom} (Normal Operative
Mode) protocol where both tasks and their subtasks are allocated samples
without any overlap of samples or clients. The data from session-I has been
used for training and validating the ResFPN by splitting in the ratio of
0.8:0.2. Thus, 508 FV presentations from session-I of SDUMLA were used to
train the ResFPN, while remaining 128 presentations were used to evaluate the
performance of the ResFPN over each training epoch. Hereafter we do not use the
presentations from first 80\% clients as these have been \textit{seen} by the
network. The remaining 20\% data is split into equal halves for \texttt{dev}
(development set) and \texttt{eval} (evaluation or test set).  In either case,
the presentations from sessions II and III are used for enrollment and those
from sessions IV, V and VI are used for probing. The protocol is summarized in
Table~\ref{tab:protocol_sdumla}. 
%
\begin{table}[]
\renewcommand{\arraystretch}{1.3}
\caption{\sc Experimental \textit{Nom} protocol for the SDUMLA database.}
\centering
\begin{tabular}{| p{3cm} | p{3cm} | p{3cm} | p{3cm} |}
\hline
Identities & \multicolumn{3}{c|}{Sessions}  \\ \cline{2-4} 
           &  I    & II / III   &  IV / V / VI  \\ \hline
1          & \cellcolor{blue!10}  & \multicolumn{2}{c|}{}    \\ 
$\vdots$   & \cellcolor{blue!10}  & \multicolumn{2}{c|}{}    \\
508 (80\%) & \multirow{-3}{*}{\cellcolor{blue!10} \thead[cc]{A \\ training \\ preprocessing}} 
           & \multicolumn{2}{c|} {\multirow{-3}{*}{unused}}  \\ \hline
572 (10\%) & \cellcolor{blue!30} & \cellcolor{orange!10} \thead{C\\ development\\ enrollment} 
           & \cellcolor{green!10} \thead{D \\ development\\ probes} \\ \cline{1-1} \cline{3-4} 
636 (10\%) & \multirow{-4}{*}{\cellcolor{blue!30} \thead[cc]{B \\ validation \\ preprocessing}}
           & \cellcolor{orange!30} \thead{E \\ evaluation\\ enrollment} 
           & \cellcolor{green!30} \thead{F \\ evaluation\\ probes}  \\ \hline
\end{tabular}
\label{tab:protocol_sdumla}
\end{table}


\begin{figure*}[ht]
\centering
\begin{subfigure}[b]{0.3\textwidth}
    \includegraphics[width=\textwidth]{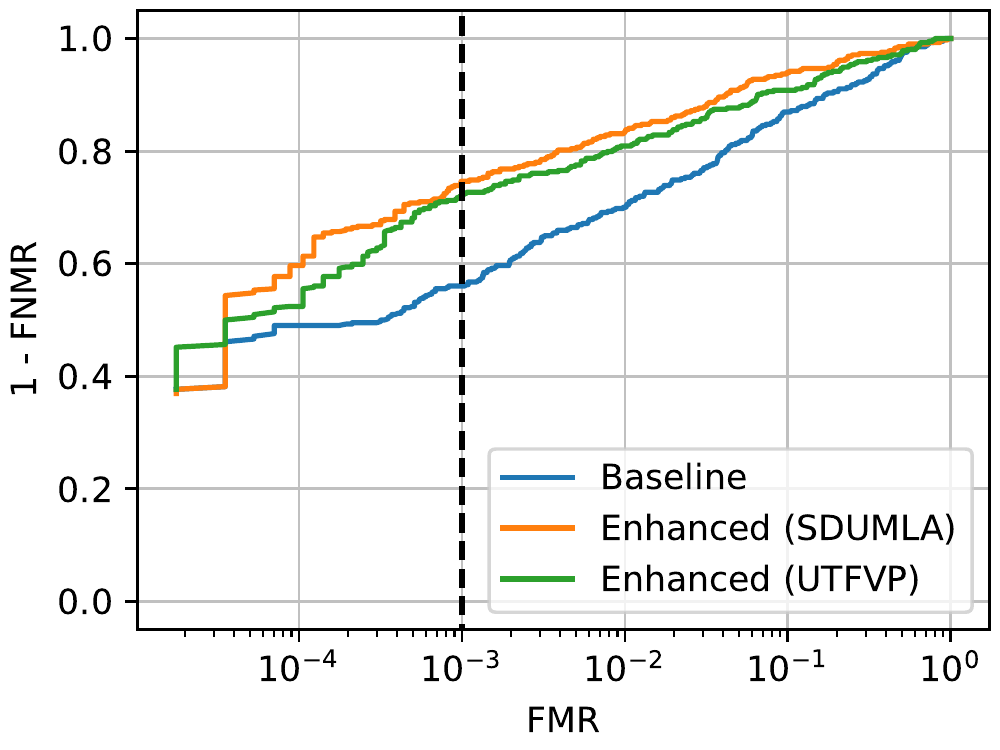}
    \caption{\texttt{dev} set}
    \label{fig:sdumla_dev}
\end{subfigure} \,
\begin{subfigure}[b]{0.3\textwidth}
    \includegraphics[width=\textwidth]{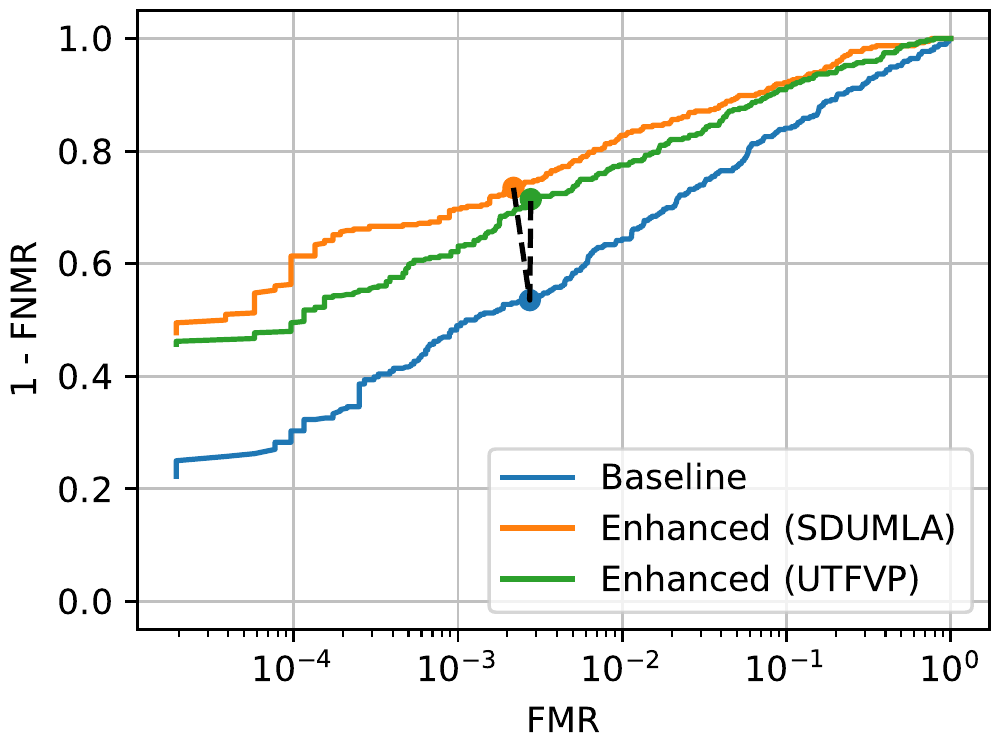}
    \caption{\texttt{eval} set}
    \label{fig:sdumla_eval}
\end{subfigure} \,
\begin{subfigure}[b]{0.35\textwidth}
    \includegraphics[width=\textwidth]{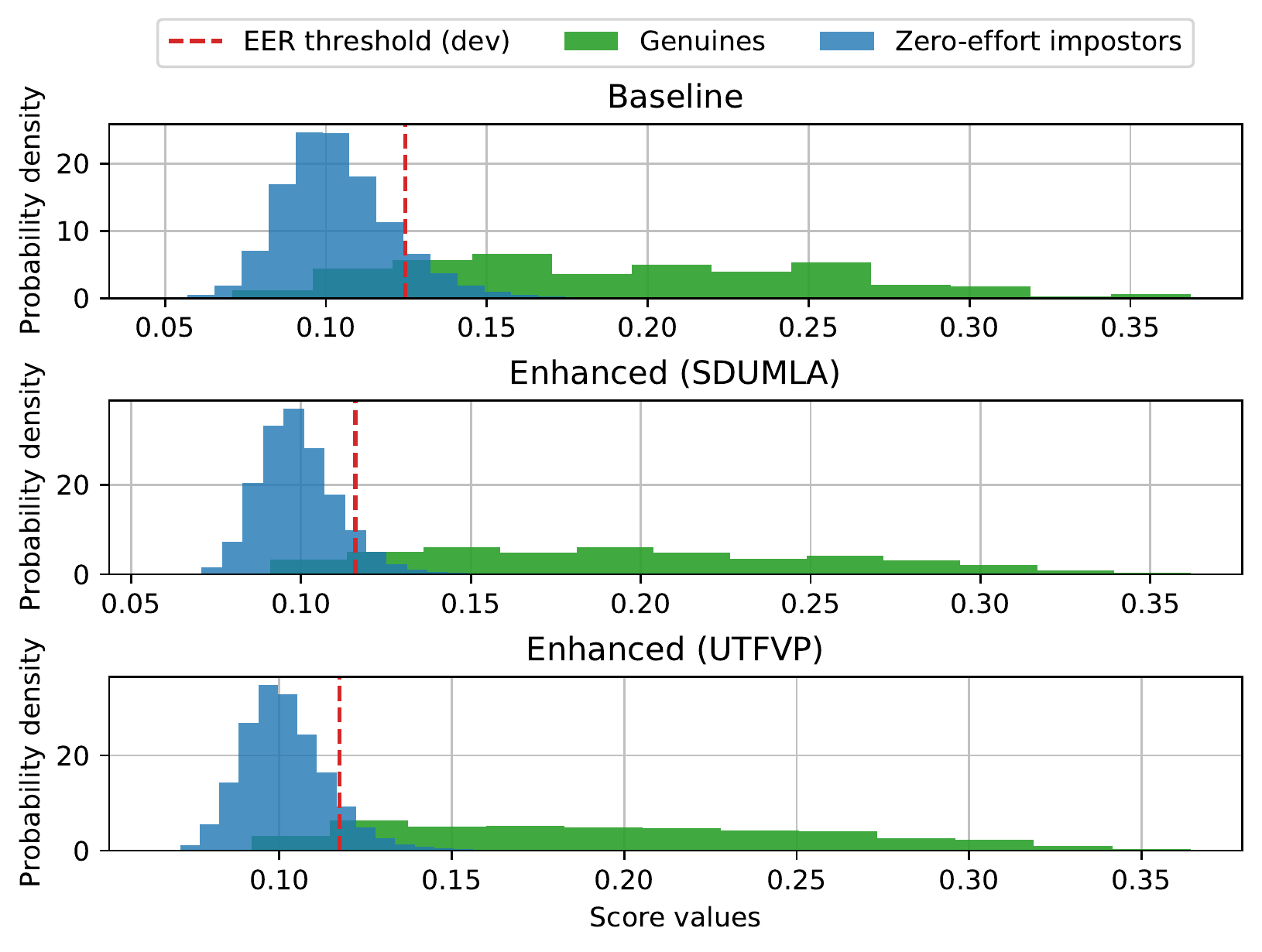}
    \caption{\texttt{eval} set}
    \label{fig:sdumla_hist}
\end{subfigure}
\caption{Receiver Operating Characteristics (ROC) curve and score histogram for the FV recognition on the Nom protocol of the SDUMLA dataset.}
\label{fig:sdumla}
\end{figure*}

The UTFVP dataset consists of 6 fingers (3 for each hand) from 60 individuals
captured twice in 2 sessions. The dataset, thus, consists of 1,440 FV
presentations with $672 \times 380$ pixels. Considering each finger as a
separate identity, we have a total of $60 \times 6 = 360$ unique fingers in the
UTFVP dataset. For experiments, we consider the \textit{Nom} (Normal Operative
Mode) protocol which is similar to the one implemented for the SDUMLA
dataset.\footnote{The details of Nom protocol as devised by Idiap Research
Institute:
\url{https://www.idiap.ch/software/bob/docs/bob/bob.db.utfvp/master}} Here, the
unique fingers from first 10 clients are considered towards training the
ResFPN. Due to small size of training set, we do not split it further for
validation, and rather the performance of model is evaluated on the training
data itself (no cross-validation for ResFPN). FV presentations from clients
11--28 constitute the \texttt{dev} set, and remaining presentations from
clients 29--60 are included in the \texttt{eval} set. We omit further details
of this protocol for the brevity of space. 


\begin{table*}[h]
\renewcommand{\arraystretch}{1.3}
\caption{\sc Performance evaluation of the proposed ResFPN for FV enhancement on the SDUMLA and UTFVP datasets along with baselines. All measure rates
are in \%. The numbers in parenthesis indicate the number of incorrectly classified samples for total samples in the given class.}
\resizebox{\textwidth}{!} {
\begin{tabular}{|l || l| l|| l| l|| l| l|} \hline
\textbf{Measure}  & \multicolumn{2}{c||}{\textbf{Baseline (No Enhancement)}} & \multicolumn{2}{c||}{\textbf{Enhanced with ResFPN (SDUMLA)}}  & \multicolumn{2}{c|}{\textbf{Enhanced with ResFPN (UTFVP)}} \\ \cline{2-7}
     &   \texttt{dev} & \texttt{eval}                  & \texttt{dev}     & \texttt{eval}         & \texttt{dev}         & \texttt{eval}           \\ \hline \hline
\multicolumn{7}{|l|}{\textbf{Test dataset: SDUMLA}}  \\ \hline
\textbf{FMR}     & 12.1 (6856/56718) & 11.4 (5922/51876) & 7.2 (4110/56718) & 8.3 (4309/51876) & 9.2 (5206/56718) & 10.4 (5401/51876) \\
\textbf{FNMR}    & 12.1 (50/414)     & 15.4 (61/396)     & 7.2 (30/414)     & 8.6 (34/396)     & 9.2 (38/414)     & 8.6 (34/396)      \\
\textbf{HTER}    & 12.1              & 13.4              & 7.2              & 8.4              & 9.2              & 9.5               \\ \hline \hline
\multicolumn{7}{|l|}{\textbf{Test dataset: UTFVP}}  \\ \hline
\textbf{FMR}     & 1.2 (274/23112)   & 1.1 (807/73344)   & 0.5 (107/23112)  & 0.3 (209/73344)  & 0.5 (107/23112)  & 0.3 (218/73344)   \\
\textbf{FNMR}    & 1.4 (3/216)       & 3.6 (14/384)      & 0.5 (1/216)      & 2.3 (9/384)      & 0.5 (1/216)      & 2.9 (11/384)      \\
\textbf{HTER}    & 1.3               & 2.4               & 0.5              & 1.3              & 0.5              & 1.6               \\ \hline
\end{tabular}
} 
\label{tab:results}
\end{table*}


\subsection{Evaluation Measures}
\label{ss:measures}
We have reported the performance of the overall FV recognition pipeline using
False Match Rate (FMR) and False Non-Match Rate (FNMR). The FMR is the ratio of
number of impostor attempts incorrectly classified as genuine matches to the
total number of impostor attempts. The FNMR is defined as the percentage of
genuine matches that are incorrectly rejected. We used the equal error rate
(EER) on the \texttt{dev} set to compute the score threshold, where FMR
$\approx$ FNMR. The Half-Total Error Rate (HTER)-- average of FMR and FNMR on
the \texttt{eval} set-- is also reported.


\subsection{Results}

\begin{figure*}[!h]
\centering
\begin{subfigure}[b]{0.3\textwidth}
    \includegraphics[width=\textwidth]{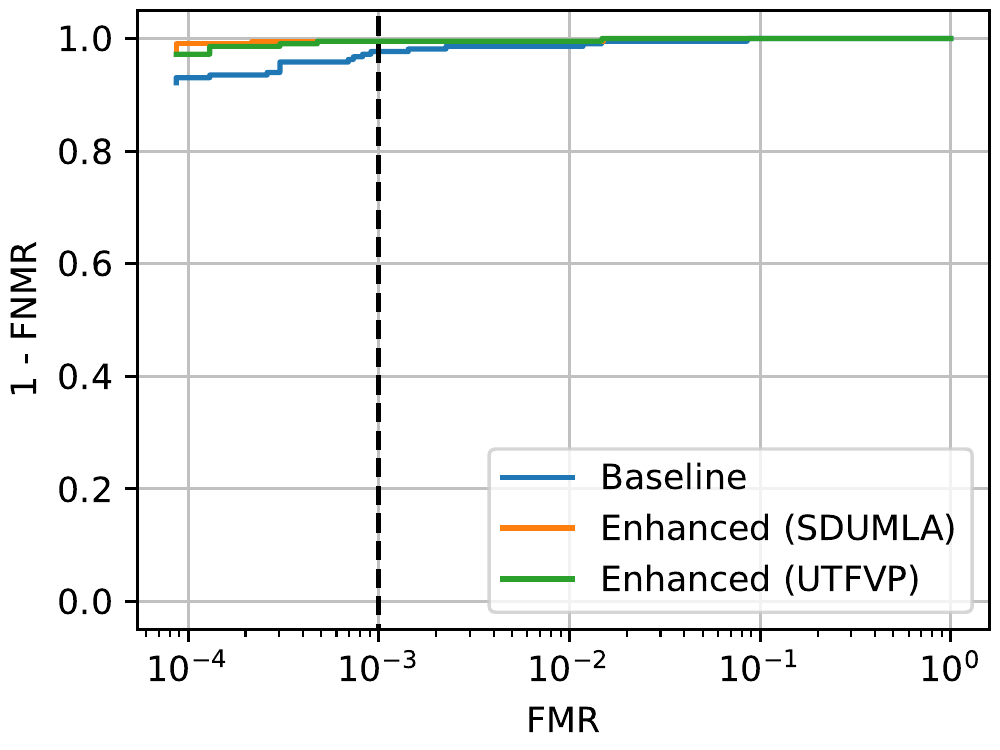}
    \caption{\texttt{dev} set}
    \label{fig:utfvp_dev}
\end{subfigure} \,
\begin{subfigure}[b]{0.3\textwidth}
    \includegraphics[width=\textwidth]{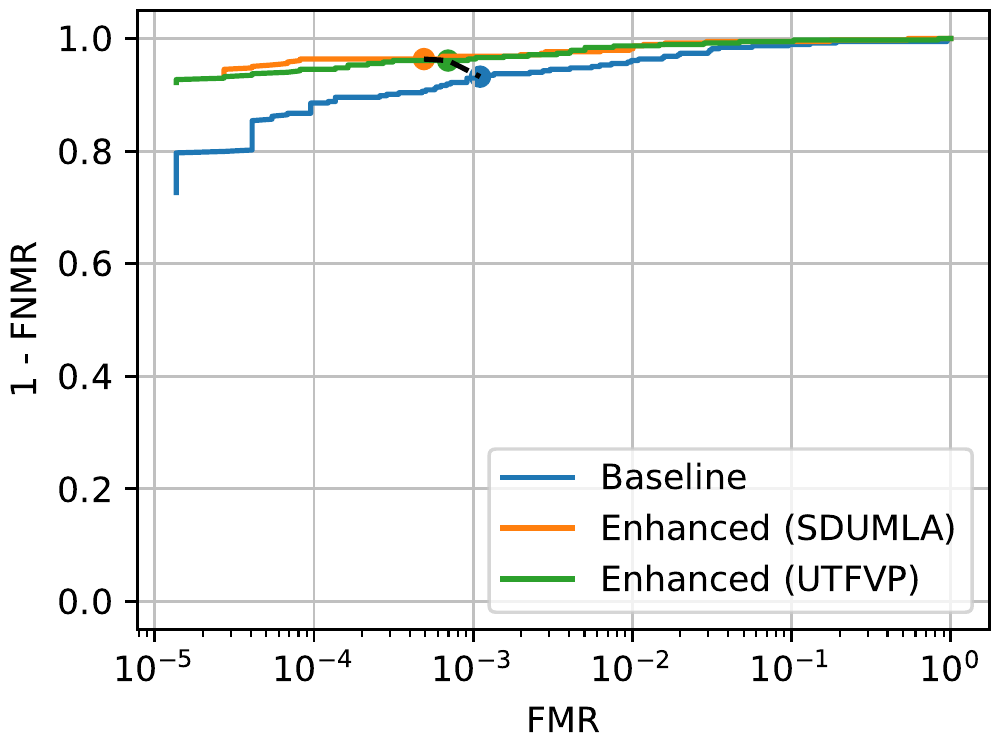}
    \caption{\texttt{eval} set}
    \label{fig:utfvp_eval}
\end{subfigure} \,
\begin{subfigure}[b]{0.35\textwidth}
    \includegraphics[width=\textwidth]{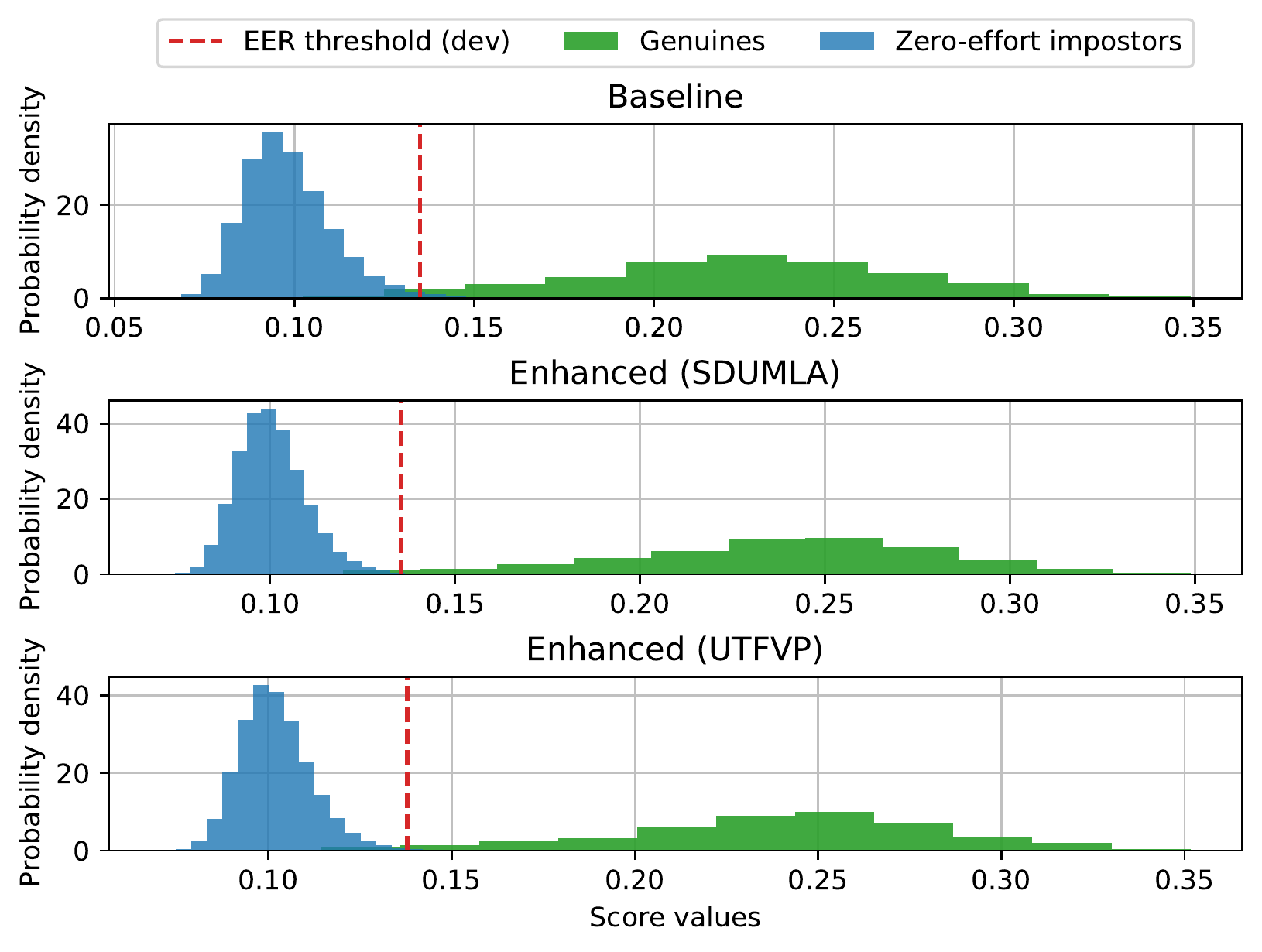}
    \caption{\texttt{eval} set}
    \label{fig:utfvp_hist}
\end{subfigure}
\caption{Receiver Operating Characteristics (ROC) curve and score histogram for the FV recognition on the Nom protocol of the UTFVP dataset.}
\label{fig:utfvp}
\end{figure*}

\textbf{Baselines:}
The recognition performances of \texttt{eval} sets of SDUMLA as well as UTFVP
datasets without applying the proposed enhancement technique are considered as
the baselines for each dataset. For SDUMLA dataset, we obtained 12.1\% EER on
its \texttt{dev} set, and 13.4\% HTER on the \texttt{eval} set. For the UTFVP
dataset, these numbers were 1.3\% and 2.4\%, respectively. The results are
summarized in Table~\ref{tab:results}. The Receiver Operating Characteristics
(ROC) plots for SDUMLA and UTFVP baselines are shown in Figures~\ref{fig:sdumla}
and \ref{fig:utfvp}, respectively (indicated by blue lines).

\textbf{Experiments on SDUMLA dataset:}
The ResFPN trained on (the train set of) SDUMLA dataset was used to enhance the
presentations of \texttt{dev} and \texttt{eval} sets of the SDUMLA dataset.
This intra-dataset experiment, however, does not have any overlapped samples or
clients across partitions. On enhanced FV presentations, we obtained the EER of
7.2\% on the \texttt{dev} set where 4,140 matches out of 57,132 were
incorrectly classified. For the \texttt{eval} set, the HTER was 8.4\% with
4,343 incorrect results out of 52,272 matches. The reduction in the overall
classification error on the \texttt{dev} as well as \texttt{eval} set is around
5\% after applying the vein-enhancement at preprocessing stage. The number of
falsely matched impostors reduced from 5,922 to 4,309 (\textit{i.e.}, nearly 27\% less)
on the \texttt{eval} set of the SDUMLA. This improvement is particularly
important since it was observed on the subset of the data that was unseen by
the ResFPN and FV recognition system. For the cross-dataset testing, we have
enhanced the FV presentations from SDUMLA using the ResFPN trained on the UTFVP
dataset.  Compared to the baseline, we observed an average improvement of 3\%
on the \texttt{dev} set for this experiment. For the \texttt{eval} set, the
number of falsely matched impostors reduced by nearly 500 samples, and the
number of incorrectly rejected genuine matches reduced to 34 from 61. In terms
of HTER, the use of vein-enhancement resulted in an improvement of 4.9\% over
the baseline. For both experiments, the performance measures are provided in
Table~\ref{tab:results} and ROC plots are shown in Figure~\ref{fig:sdumla}.  It
may be observed from the ROCs that the performance of the FV recognition using
enhanced presentations is consistently better than the baseline (without
enhancement) over a complete range of FMR. This relative improvement is
highlighted even more on the ROC of \texttt{eval} set at lower values of FMR.
For enhanced presentations, the score histograms of the \texttt{eval} set
(Figure~\ref{fig:sdumla_hist}) indicate a better separation between scores of
both classes, and also lowered mean and lesser variance of the scores of the
impostor comparisons.

\textbf{Experiments on UTFVP dataset:} 
The Nom protocol of the UTFVP dataset comprises 73,344 impostor comparisons and
384 genuine comparisons on the \texttt{eval} set. When the presentations were
enhanced using the proposed ResFPN (trained on the SDUMLA, \textit{i.e.} cross-dataset),
209 impostor comparisons were incorrectly classified as genuine. This number is
approximately 1/4-th of the same metric obtained for non-enhanced version of
same presentations. On the \texttt{dev} set, we observed 40\% reduction for
this metric with respect to the baseline. The FNMR and, thus, HTER on both sets
of the UTFVP dataset also improved by 0.8--1.3\% when the performance of FV
recognition was evaluated in the cross-dataset scenario as detailed in
Table~\ref{tab:results}. 

Interestingly, when the FV presentations were enhanced using ResFPN trained on
the other (disjoint) partition of UTFVP, we observed the improvement, in terms
of FMR/FNMR, to be similar to the aforementioned cross-dataset experiment.  The
total number of misclassifications on the \texttt{dev} set of the UTFVP reduced
from 277 (in baseline) to 108 for both experiments of vein enhancement. This
improvement was even better for the \texttt{eval} set where misclassifications
reduced from 821 to 218--229 after enhancement of the input. While it may
appear that the ResFPN trained on the subset of UTFVP has performed relatively
poorer than the network trained on the SDUMLA dataset, it may be noted that the
\texttt{train} set of UTFVP consists of only 388 presentations, which is much
smaller than its SDUMLA counterpart. 

The ROC plots for the \texttt{dev} set are near-perfect for both enhancement
experiments as indicated by almost horizontal curves in
Figure~\ref{fig:utfvp_dev}.  While the performance of baseline experiment
slowly degrades for FMR $ < 10^{-3}$, the recognition of the enhanced
presentations remains consistently accurate. On the \texttt{eval} set, one can
observe that the improvement in FV recognition, brought by the ResFPN, is
similar for models trained on SDUMLA as well as UTFVP.
Figure~\ref{fig:utfvp_hist} shows the overall increase in the genuine scores of
the enhanced presentations which improves separability of genuine comparisons
from the impostor attempts.


\section{Conclusions}
\label{sec:conc}
In this work, we have proposed a ResFPN (Residual Feature Pyramid Network) for
enhancement of vascular patterns in the FV presentations acquired in NIR. This
network can be integrated into a standard recognition pipeline as a part of
preprocessing module. With its peculiar SDBlock and FAM architectures, the
proposed network is able to detect vein-structures at various scales and
combine them efficiently to generate an enhanced presentation.  With usage of
enhanced data, the performance of recognition system has improved in terms of
FMR, FNMR, and HTER-- over different datasets as demonstrated by our results.
Thus, the resultant recognition systems are more accurate and secure. 

We have introduced a novel network architecture for detection of
vein-structures. Further work in this direction mainly includes better
generalization across variety of recognition methods, and efficiently
processing size-independent presentations.


\section*{Acknowledgement}
The authors would like to thank InnoSuisse for the project CANDY, and 
the Swiss Center for Biometrics Research and Testing for their support.


\bibliographystyle{ieeetr}
\bibliography{egbib}

\end{document}